\documentclass[3p,times,twocolumn]{elsarticle}
\usepackage{graphicx}
\usepackage{here}
\usepackage{ecrc}

\volume{00}

\firstpage{1}

\journalname{Nuclear and Particle Physics Proceedings}

\runauth{}

\jid{nppp}

\jnltitlelogo{Nuclear and Particle Physics Proceedings}

\usepackage{amssymb}
\usepackage[figuresright]{rotating}
\newcommand{\be}{\begin{equation}}

\newcommand{\ee}{\end{equation}}
\newcommand{\bea}{\begin{eqnarray}}
\newcommand{\eea}{\end{eqnarray}}
\newcommand{\beas}{\begin{eqnarray*}}
\newcommand{\eeas}{\end{eqnarray*}}

\newcommand{\nn}{\nonumber\\}
\newcommand{\slsh}[1]{{\not \! #1}}
\newcommand{\slshh}[1]{{\not \!\! #1}}

\begin{document}

\begin{frontmatter}

%%
%%%%%%%%%%%%%%%%%%%%%%%%%%%%%%%%%%%%%%%%%%%%%%%%%
\title{Thermo-magnetic behavior of the quark-gluon vertex
 $^*$}
 % \corref{cor0}}
 \cortext[cor0]{Talk given at 18th International Conference in Quantum Chromodynamics (QCD 15,  30th anniversary),  29 june - 3 july 2015, Montpellier - FR}
 \author[label1,label4]{Alejandro Ayala}
%  \cortext[cor0]{FAPESP CNPq-Brasil PhD student fellow.}
\ead{ayala@nucleares.unam.mx}
\address[label1]{Instituto de Ciencias
  Nucleares, Universidad Nacional Aut\'onoma de M\'exico, Apartado
  Postal 70-543, M\'exico Distrito Federal 04510,
  Mexico.}
\address[label4]{Centre for Theoretical and Mathematical Physics, and Department of Physics,
  University of Cape Town, Rondebosch 7700, South Africa.}
 \author[label2]{J. J. Cobos-Mart\'{\i}nez}
  \ead{j.j.cobos.martinez@gmail.com}
\address[label2]{Instituto de F\'{\i}sica y Matem\'aticas, Universidad Michoacana de San Nicol\'as de Hidalgo,
Edificio C-3, Ciudad Universitaria, Morelia, Michoac\'an 58040, M\'exico.}
 \author[label3,label4,label5]{M. Loewe\fnref{fn1}}
   \fntext[fn1]{Speaker, Corresponding author.}
    \ead{mloewe@fis.puc.cl}
    \address[label3]{Instituto de F\1sica, Pontificia Universidad Cat\'olica de Chile,
  Casilla 306, Santiago 22, Chile.}
   \address[label5]{Centro Cient\'{\i}fico-Tecnol\'ogico de Valpara\'{\i}so, Casilla 110-V, Valpara\'{\i}so, Chile.}
 \author[label6,label1]{Mar\'{\i}a Elena Tejeda-Yeomans}
  \ead{elena.tejeda@fisica.uson.mx}
   \address[label6]{Departamento de F\'{\i}sica, Universidad de Sonora, Boulevard Luis Encinas y Rosales, Colonia Centro, Hermosilla, Sonora 83000, M\'exico}
   \author[label3]{R. Zamora}
   \ead{rrzamora@uc.cl}

\pagestyle{myheadings}
\markright{ }
\begin{abstract}
The thermo-magnetic corrections to the quark-gluon vertex in the presence of a weak magnetic field are calculated  in the frame of  the Hard Thermal Loop approximation. The vertex satisfies a QED-like Ward identity with the quark self-energy calculated within  the same approximation. It turns out that only the longitudinal vertex components get modified. The calculation provides a first principles result for the quark anomalous magnetic moment at high temperature in a weak magnetic field.  The effective thermo-magnetic quark-gluon coupling shows a decreasing behavior as function of the field strength. This result supports the observation that the behavior  of the effective quark-gluon coupling in the presence of a magnetic field is an important ingredient in order to understand the inverse magnetic catalysis phenomenon recently observed in the lattice QCD simulations.
\end{abstract}
% \begin{document}

\end{frontmatter}
%%%%%%%%%%%%
%\vspace*{-1.5cm}
\section{Introduction}\label{Introduction}
\noindent
 Recent lattice QCD results, with 2+1 quark flavors,  indicate that the transition temperature  measured from the behavior of the chiral condensate and susceptibility, as well as from other thermodynamic observables such as longitudinal and transverse pressure, magnetization and energy and entropy densities, decreases with increasing magnetic field~\cite{Fodor,Bali:2012zg,Bali2}. This result was called  {\it magnetic anticatalysis}.

\noindent
Recently, we  have shown, in a natural way, how  the decrease of the coupling constant with increasing field strength can be obtained within a perturbative calculation beyond the mean field approximation. Notice that such a behavior cannot be achived in the mean field approximation ~\cite{mcz}. This was done both for the abelian Higgs ~\cite{amlz}  model as well as for  the linear sigma model ~\cite{alz}  where charged fields are subject to the effect of a constant magnetic field. This behavior introduces a dependence of the boson masses on the magnetic field inducing then a decreasing behavior  of the critical temperature for chiral symmetry breaking/restoration. Going now into QCD,  in order to to establish if a  similar behavior takes place, a first step would be to determine the finite temperature and magnetic field dependence of the coupling constant.

\noindent
The discussion of the thermal behavior of systems involving massless bosons, such as gluons, in the presence of magnetic fields is quite subtle. Unless a careful treatment is implemented, severe  infrared divergences associated to the effective dimensional reduction of the momentum integrals will  appear. These divergencs are associated  to  the separation of the energy levels into transverse and longitudinal directions (with respect to the magnetic field direction). The former are given in terms of discrete Landau levels. Thus, the longitudinal mode alone no longer can tame the divergence of the Bose-Einstein distribution. In this context,  recently it has  been shown that it is possible to find the appropriate condensation conditions by accounting for the plasma screening effects~\cite{Ayala2}.  Here we show that a simple prescription where the fermion mass acts as the infrared regulator allows to obtain the leading behavior of the QCD coupling for weak magnetic fields at high temperature, that is, in the Hard Thermal Loop (HTL) approximation. As a first step, we compute the thermo-magnetic corrections to the quark-gluon vertex in the weak field approximation. We use this calculation to compute the thermo-magnetic dependence of the QCD coupling. To include the magnetic field effects we use Schwinger's proper time method. We should point out that the weak field approximation means that one considers the field strength to be smaller than the square of the temperature but does not imply a hierarchy with respect to other scales in the problem such as the fermion mass.

\section{Charged fermion propagator in a medium}\label{II}

\noindent
The presence of a constant magnetic field breaks Lorentz invariance and leads to a charged fermion propagator which is function of the separate transverse and longitudinal momentum components (with respect to the field direction). Considering the case of a magnetic field pointing along the $\hat{z}$ direction, namely $\vec{B}=B\hat{z}$, the vector potential, in the so called {\it symmetric gauge}, is
\bea
   A_\mu(x)=\frac{B}{2}(0,-x_2,x_1,0).
\label{symgauge}
\eea
The fermion propagator in coordinate space cannot longer be written as a simple Fourier transform of a momentum propagator but instead it is written as~\cite{Schwinger}
\bea
   S(x,x')=\Phi (x,x')\int\frac{d^4p}{(2\pi)^4}e^{-ip\cdot (x-x')}S(k),
\label{genprop}
\eea
where

\bea
  \Phi (x,x')=\exp\left\{iq\int_{x'}^xd\xi^\mu\left[A_\mu + \frac{1}{2}F_{\mu\nu}(\xi - x')^\nu\right]\right\},
\label{phase}
\eea
is called the {\it phase factor} and $q$ is the absolute value of the fermion's charge, in units of the electron charge. $S(k)$ is given by
\bea
   S(k)&=&-i\int_0^\infty \frac{ds}{\cos (qBs)} e^{is(k_{\|}^2-k_\perp^2
   \frac{\tan (qBs)}{qBs} - m^2)}\nn
   &\times& \Big\{\left[\cos (qBs) + \gamma_1 \gamma_2 \sin (qBs) \right] (m+\slsh{k_{\|}}) \nn
   &-& \frac{\slsh{k_\bot}}{\cos(qBs)} \Big\},
\label{Schwinger}
\eea
where $m$ is the quark mass and we use the definitions for the parallel and perpendicular components of the scalar product of two vectors $a^\mu$ and $b^\mu$ given by
\bea
   (a\cdot b)_{\|} &=& a_0b_0 - a_3b_3\nn
   (a\cdot b)_{\bot} &=& a_1b_1 + a_2b_2.
\label{defs}
\eea
%\begin{widetext}
%%%%%%%%%%%%%%%%%%%%%%%%%%%%%%%%%%%
\begin{center}
\begin{figure}[hbt]
\includegraphics[width=0.2\textwidth]{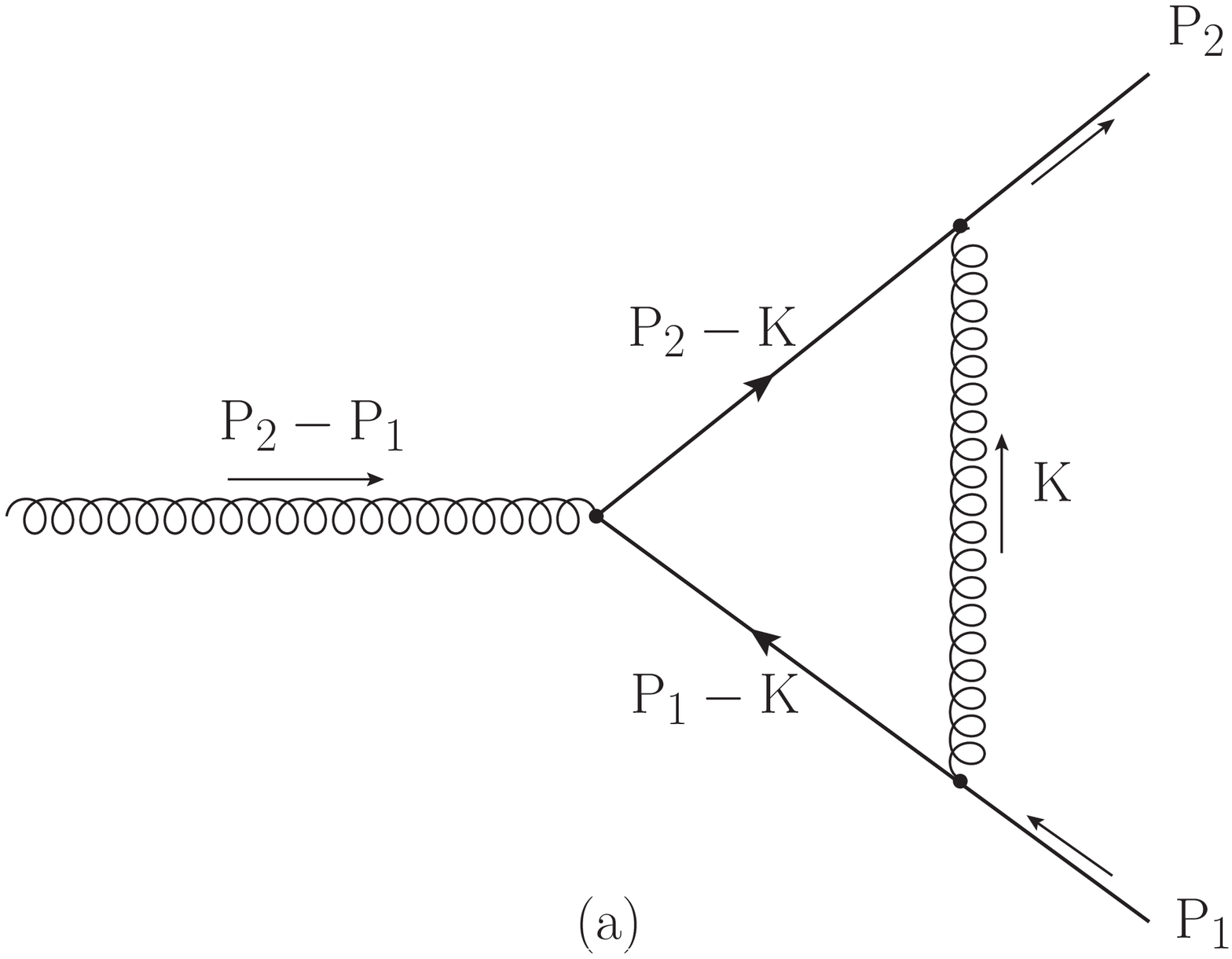}
\includegraphics[width=0.2\textwidth]{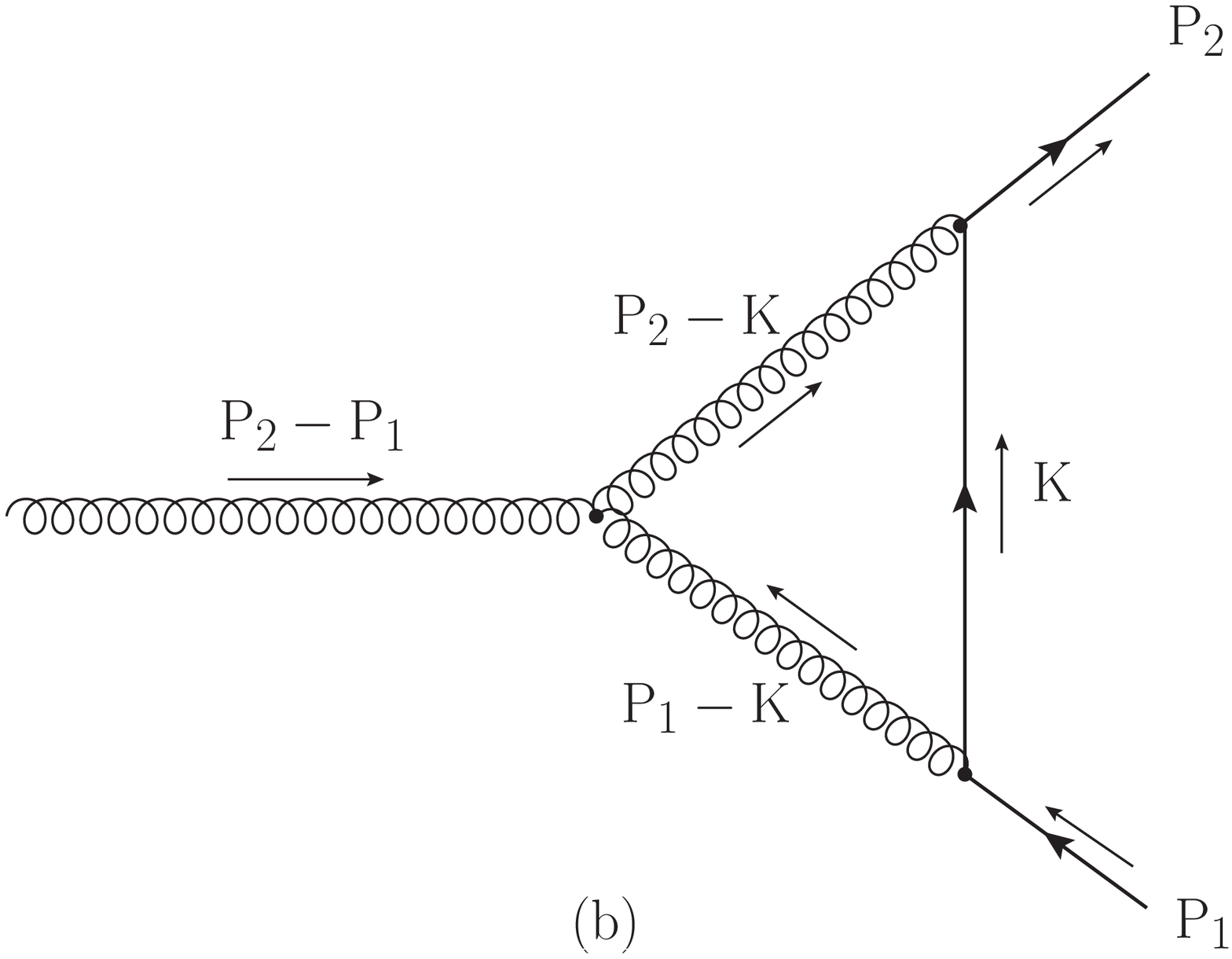}
\caption{Feynman diagrams contributing to the thermo-magnetic dependence of the quark-gluon vertex. Diagram (a) corresponds to a  QED-like contribution whereas diagram (b) corresponds to a pure QCD contribution.} \label{fig1}
\end{figure}
\end{center}

%%%%%%%%%%%%%%%%%%%%%%%%%%%%%%%%%%%
%\end{widetext}
\noindent
Figure~\ref{fig1} shows the Feynman diagrams contributing to the quark-gluon vertex. Diagram (a) corresponds to a QED-like contribution whereas diagram (b) corresponds to a pure QCD contribution. The computation of these diagrams requires using the fermion propagator given by Eq.~(\ref{genprop}), which involves the phase factor in Eq.~(\ref{phase}).  It turns out, however, that this phase does not contribute. We refer the reader to the original article ~\cite{original} where a proof of this fact is presented. Therefore, for the computation of diagrams (a) and (b) in Fig.~\ref{fig1}, we can just work with the momentum representation of the fermion propagators since the phase factors do not contribute. The situation would have been nontrivial in case the computation had required a three fermion propagator closed loop~\cite{Chyi}.

\section{QCD vertex at finite temperature with a weak magnetic field}\label{III}

\noindent
To compute the leading magnetic field dependence of the vertex at high temperature, we work in the weak field limit of the momentum representation of the fermion propagator~\cite{Chyi}. We work in Euclidean space which is suited for calculations at finite temperature in the imaginary-time formalism of finite temperature field theory. The fermion propagator up to ${\mathcal{O}}(qB)$ is written as
\bea
   S(K)=\frac{m-\slshh{K}}{K^2+m^2} - i\gamma_1\gamma_2\frac{m-\slshh{K_{\|}}}{(K^2+m^2)^2}(qB).
\label{OqB}
\eea

\noindent
Using the propagator in Eq.~(\ref{OqB}) and extracting a factor $gt_a$ common to the bare and purely thermal contributions to the vertex, the magnetic field dependent part of diagram (a) in Fig.~\ref{fig1} is expressed as
\bea
   \delta\Gamma_\mu^{\mbox{(a)}}&=&-ig^2(C_F - C_A/2)(qB)T\sum_n\int\frac{d^3k}{(2\pi)^3}\nn
   &\times&\gamma_\nu
   \left[ \gamma_1\gamma_2\slshh{K_{\|}}\gamma_\mu\slshh{K}\widetilde{\Delta}(P_2-K)\right.\nn
   &+& \left.
           \slshh{K}\gamma_\mu\gamma_1\gamma_2\slshh{K_{\|}}\widetilde{\Delta}(P_1-K)
   \right]\gamma_\nu\nn
   &\times&\Delta(K)\widetilde{\Delta}(P_2-K)\widetilde{\Delta}(P_1-K),
\label{(a)}
\eea
where, in the spirit of the HTL approximation, we have ignored terms proportional to $m$ in the numerator and
\bea
   \widetilde{\Delta}(K)&\equiv&\frac{1}{\widetilde{\omega}_n^2 + k^2 + m^2}\nn
   \Delta(K)&\equiv&\frac{1}{\omega_n^2 + k^2}
\label{Deltas}
\eea
with $\widetilde{\omega}_n=(2n+1)\pi T$ and $\omega_n=2n\pi T$ the fermion and boson Matsubara frequencies. $C_F$, $C_A$ are the factors corresponding to the fundamental and adjoint representations of the $SU(N)$ Casimir operators
\bea
   C_F&=&\frac{N^2-1}{2N}\nn
   C_A&=&N,
\label{Casimir}
\eea
respectively. Hereafter, capital letters are used to refer to four-momenta in Euclidean space with components $K_\mu=(k_4,\vec{k})=(-\omega,\vec{k})$, with $\omega$ either a Matsubara fermion or boson frequency.

\noindent Following the same procedure, the magnetic field dependent part of diagram (b) in Fig.~\ref{fig1} is expressed as
\bea
   \delta\Gamma_\mu^{\mbox{(b)}}&=&-2ig^2\frac{C_A}{2}(qB)T\sum_n\int\frac{d^3k}{(2\pi)^3}\nn
   &\times&\left[-\slshh{K}\gamma_1\gamma_2\slshh{K_{\|}}\gamma_\mu
                         +2\gamma_\nu\gamma_1\gamma_2\slshh{K_{\|}}\gamma_\nu K_\mu\right.\nn
   &-&\left.\gamma_\mu\gamma_1\gamma_2\slshh{K_{\|}}\slshh{K}
   \right]\nn
   &\times&\widetilde{\Delta}(K)^2\Delta (P_1-K)\Delta (P_2-K).
\label{(b)}
\eea

\noindent
The explicit factor $2$ on the right-hand side of Eq.~(\ref{(b)}) accounts for the two possible fermion channels. These two channels are already accounted for in Eq.~(\ref{(a)}) since the magnetic field insertion on each quark internal line is thereby included at the order we are considering. 

\noindent These expressions can be simplified by noting that in the HTL approximation the external momenta $P_{1}$ and $P_{2}$ are small and can be taken of the same order.  We have carried out the calculation of the Matsubara sums but we do not have space here to present all the details.

\noindent Adding both contributions to the vertex corrections we find a quite compact expression

\bea
   \delta\Gamma_\mu&=&\delta\Gamma_\mu^{\mbox{(a)}}+\delta\Gamma_\mu^{\mbox{(b)}}\nn
   &=&2i\gamma_5g^2C_F(qB)G_\mu(P_1,P_2),
\label{deltaGamma}
\eea

\noindent where

\bea
  G_\mu(P_1,P_2)&=&2T\sum_n\int\frac{d^3k}{(2\pi)^3}\nn
  &\times& \Big\{
  (K\cdot b)\slshh{K}u_\mu - (K\cdot u)\slshh{K}b_\mu\nn
  &+& 
  \left[ (K\cdot b)\slsh{u} - (K\cdot u)\slsh{b}\right]K_\mu
  \Big\}\nn
  &\times&\widetilde{\Delta}^2(K)\Delta(P_1-K)\Delta(P_2-K).
\label{Gmu}
\eea

\begin{center}
\begin{figure}[hbt]
\includegraphics[width=0.25\textwidth]{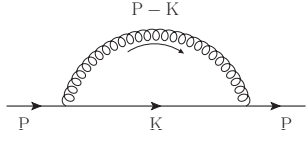}
\caption{Feynman diagram for the the quark self-energy. The internal quark line represents the quark propagator in the presence of the magnetic field in the weak field limit.}\label{fig2}
\end{figure}
\end{center}

 \noindent
 It is no surprise that the thermo-magnetic correction to the quark-gluon vertex is proportional to $\gamma_5$ since the magnetic field is odd under parity conjugation. It is important to stress  that our vertex satisfies a QED-like Ward identity, i.e.

\bea
   (P_1-P_2)\cdot\delta\Gamma (P_1,P_2)= \Sigma (P_1) - \Sigma (P_2),
\label{W2}
\eea
\noindent where $\Sigma(P)$ denotes the self energy correction for the quark in the presence of an external magnetic field. This can be computed from the diagram in Fig.~\ref{fig2} and the corresponding expression in the HTL-approximation is given by

\bea
   \Sigma(P)  = 2i\gamma_5g^2C_F(qB)
    T\sum_n 
    \int\frac{d^3k}{(2\pi)^3} \nn \Delta(P-K)\widetilde{\Delta}^2(K)
  \times 
  \Big\{ (K\cdot b)\slsh{u} - (K\cdot u)\slsh{b}\Big\}.
 \label{selfenergy}
\eea

\noindent  The technical details requeried for computing the frequency sums cannot be presented here due to lack of space. The reader is refered once again to  ~\cite{original}.

\noindent
This remarkable result shows that even in the presence of the magnetic field and provided the temperature is the largest of the energy scales, the thermo-magnetic correction to the quark-gluon vertex is gauge invariant.

\section{Thermo-magnetic QCD coupling}\label{IV}

\noindent The calculation of $G_\mu (P_{1}, P_{2})$ demands not only to deal with Matsubara sums but also to discuss tensor expressions of the type

\bea
   J_{\alpha i}(P_1,P_2)\equiv\int\frac{d\Omega}{4\pi}\frac{\hat{K}_\alpha\hat{K}_i}
   {(P_1\cdot{\hat{K}})(P_2\cdot{\hat{K}})}.
\label{Js}
\eea

\noindent For the sake of simplicity we have chosen a particular configuration where the external momenta $\vec{P}_{1}$ and $\vec{P}_{2}$ make a relative angle $\theta _{12} = \pi$. It turns out that only the longitudinal components of our tensor,  i.e of the thermomagnetic vertex, are modified. If we rearrange the gamma matrices to introduce the spin operator in the $\hat{z}$ direction $\Sigma _{3}$, we find
for the explicit longitudinal components 
\bea
  \vec{\delta\Gamma}_\parallel(p_0) = \left(\frac{2}{3p_0^2}\right)4g^2C_FM^2(T,m,qB)\vec{\gamma}_\parallel\Sigma_3,
\label{components}
\eea
where $\vec{\gamma}_\parallel = (\gamma_0,0,0,-\gamma_3) $ and where we have defined the function $M^2(T,m,qB)$ as
\bea
   M^2(T,m,qB)=\frac{qB}{16\pi^2}\left[\ln(2) - \frac{\pi}{2}\frac{T}{m}\right].
\label{M}
\eea 

\noindent
That  the vertex correction in Eq.~(\ref{components}) is proportional to the third component of the spin operator is of course natural since the first order magnetic correction to the vertex is in turn proportional to the spin interaction with the magnetic field, which we chose to point along the third spatial direction. The correction, thus,  corresponds to the quark anomalous magnetic moment at high temperature in a weak magnetic field. Note also that Eq.~(\ref{components}) depends on the scales $p_0$ and $m$. $p_0$ is the typical energy of a quark in the medium and therefore the simplest choice for this scale is to take it as the temperature. The quark mass represents the infrared scale and it is therefore natural to take it as the thermal quark mass. We thus set
\bea
   p_0&=&T\nn
   m^2&=&m_f^2=\frac{1}{8}g^2T^2C_F.
\label{choise}
\eea

\noindent The pure thermal correction to the quark-gluon vertex is well known ~\cite{LeBellac} to be given by

\bea
   \delta\Gamma^{\mbox{\small{therm}}}_\mu(P_1,P_2)=-m_f^2\int\frac{d\Omega}{4\pi}
   \frac{\hat{K}_\mu\slshh{\hat{K}}}{(P_1\cdot\hat{K})(P_2\cdot\hat{K})}.
\label{purether}
\eea
In order to extract the effective modification to the coupling constant in one of the longitudinal directions, let us look at the contribution proportional to $\gamma_0$ from Eq.~(\ref{purether}). For the same working momenta configuration we find
\bea
   \delta\Gamma^{\mbox{\small{therm}}}_0(p_0)=-\frac{m_f^2}{p_0^2}\gamma_0.
\label{purether0}
\eea
Using Eqs.~(\ref{components}) and~(\ref{purether0}), the effective thermo-magnetic modification to the quark-gluon coupling extracted from the effective longitudinal vertex, is given by
\bea
   g_{\mbox{\small{eff}}}=g\left[1-\frac{m_f^2}{T^2}+\left(\frac{8}{3T^2}\right)g^2C_FM^2(T,m_f,qB)\right],
\label{effective}
\eea

\begin{center}
\begin{figure}[hbt]
\includegraphics[width=0.50\textwidth]{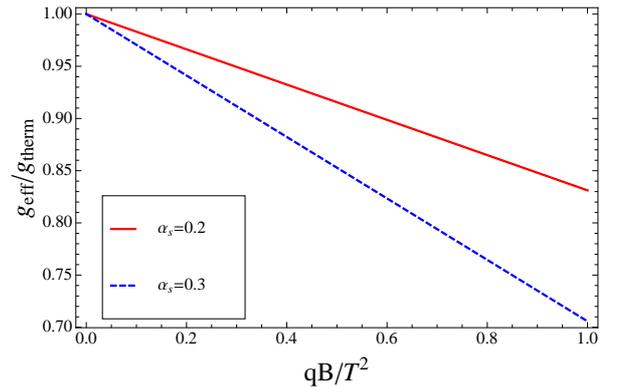}
\caption{The effective thermo-magnetic coupling $g_{\mbox{\small{eff}}}$ normalized to the purely thermal coupling $g_{\mbox{\small{therm}}}$ as a function of the field strength scaled by the squared of the temperature, for $\alpha_s=0.2, 0.3$. Note that $g_{\mbox{\small{eff}}}$ decreases down to about 15\% -- 25\% for the largest strength of the magnetic field within the weak field limit with respect to the purely thermal correction and that the decrease is faster for larger $\alpha_s$.}
\label{fig4}
\end{figure}
\end{center}

\noindent Figure~\ref{fig4} shows the behavior of $g_{\mbox{\small{eff}}}$ normalized to $g_{\mbox{\small{therm}}}$, where
\bea
   g_{\mbox{\small{therm}}}\equiv g\left(1-\frac{m_f^2}{T^2}\right),
\label{effectivetherm}
\eea
for $\alpha_s=g^2/4\pi =0.2, 0.3$ as a function of the scaled variable $b=qB/T^2$. Plotted as a function of $b$ and for our choice of momentum scales, the function $g_{\mbox{\small{eff}}}/g_{\mbox{\small{therm}}}$ is temperature independent. 

\noindent Note that the effective thermo-magnetic coupling $g_{\mbox{\small{eff}}}$ is a decreasing function of the magnetic field. The decrease becomes more significant for larger $\alpha_s$ and for the considered values of $\alpha_s$ it becomes about 15\% -- 25\% smaller than the purely thermal correction for $qB\sim T^2\sim 1$.

\section{Summary and Conclusions}\label{conclusions}

\noindent
In this work we have computed the thermo-magnetic corrections to the quark-gluon vertex for a weak magnetic field and in the HTL approximation. It turns out that this vertex satisfies a QED-like Ward identity with the quark self-energy. This results hints to the gauge-independence of the calculation. The thermo-magnetic correction is proportional to the spin component in the direction of the magnetic field, affecting only the longitudinal components of the quark-gluon vertex. It thus corresponds to the quark anomalous magnetic moment at high temperature in a weak magnetic field. 

\noindent
From the quark-gluon vertex we  extracted the behavior of the magnetic field dependence of the QCD coupling. The coupling decreases as the field strength increases. For analytic simplicity, the explicit calculation has been performed for a momentum configuration that accounts for the conditions prevailing in a quark-gluon medium at high temperature, namely namely back-to-back slow moving quarks whose energy is of the order of the temperature and with the infrared scale of the order of the thermal particle's mass. Notice that the chosen values for these scales provide a good representation of the coupling constant's strength within the plasma conditions.

\noindent
We stress that the weak field approximation means that the field strength is smaller than the square of the temperature but does not require a hierarchy with respect to other scales in the problem such as the fermion thermal mass. Note that for situations where the temperature is not the largest of the energy scales, as for the largest magnetic field strengths achieved in peripheral heavy-ion collisions, or for compact astrophysical objects, the present calculation does not provide conclusions about the behavior of the coupling constant nor for its effect on the critical temperature. Nevertheless, it is important to note that lattice QCD simulations for the critical temperature are performed for magnetic field intensities starting from $qB=0$ (see {\it e.g.} Fig. 9 in Ref.~\cite{Fodor}). The results of these simulations show that the critical temperature is a decreasing function of the field intensity all the way to $qB= 1$ GeV$^2$. Also, one should note that it is likely that the effects on the quark-gluon plasma in peripheral heavy-ion collisions take place when the field intensity is already smaller than the square of the temperature, since the field strength is a rapidly decreasing function of time. It is thus important to explore in QCD the regime where $qB$ is small compared to $T^2$ as a necessary bridge to further studies on the behavior of the coupling constant for larger field strengths. 

\noindent
The result supports the idea~\cite{amlz, alz} that the decreasing of the coupling constant may be one of the important ingredients to understand the inverse magnetic catalysis obtained in lattice QCD.

\section*{Acknowledgments}

A. A. acknowledges useful conversations with G. Krein. J. J. C.-M. acknowledges support from a CONACyT-M\'exico post\-doc\-to\-ral grant with num\-ber 290807-UMSNH. Support for this work has been received in part from CONACyT-M\'exico under grant number 128534 and FONDECYT under grant  numbers 1130056 and 1120770. R. Z. acknowledges support from CONICYT under Grant No. 21110295.

\end{document}